\documentclass[pre,showpacs,superscriptaddress,twocolumn]{revtex4-1}
\usepackage{amsmath}
\usepackage{amsfonts}
\usepackage{amssymb}
\usepackage{graphicx}
\usepackage{bm}
\usepackage{hyperref}

\usepackage{color}
\usepackage{array,tabularx}
\usepackage{epstopdf}

\begin{document}
\title{Long-wavelength fluctuations and anomalous dynamics in two-dimensional liquids}

\author{Yan-Wei Li}
\affiliation{Division of Physics and Applied Physics, School of Physical and
Mathematical Sciences, Nanyang Technological University, Singapore}
\author{Chandan K. Mishra}
\affiliation{Chemistry and Physics of Materials Unit, Jawaharlal Nehru Centre for Advanced Scientific Research, Jakkur, Bangalore 560064, India}
\affiliation{Department of Physics and Astronomy, University of Pennsylvania, Philadelphia, PA, USA}
\author{Zhaoyan Sun}
\affiliation{State Key Laboratory of Polymer Physics and Chemistry, Changchun
Institute of Applied Chemistry, Chinese Academy of Sciences, Changchun 130022,
China}
\affiliation{University of Science and Technology of China, Hefei, 230026,
China}
\author{Kun Zhao}
\affiliation{Key Laboratory of Systems Bioengineering (Ministry of Education),
School of Chemical Engineering and Technology, Tianjin University, Tianjin
300072, China}
\author{Thomas G. Mason}
\affiliation{Department of Chemistry and Biochemistry, University of California,
Los Angeles, CA 90095 USA}
\affiliation{Department of Physics and Astronomy, University of California, Los
Angeles, CA 90095 USA}
\author{Rajesh Ganapathy}
\affiliation{International Centre for Materials Science, Jawaharlal Nehru Centre for Advanced Scientific Research,
Jakkur, Bangalore 560064, India}
\author{Massimo Pica Ciamarra}
\email{massimo@ntu.edu.sg}
\affiliation{Division of Physics and Applied Physics, School of Physical and
Mathematical Sciences, Nanyang Technological University, Singapore}
\affiliation{
CNR--SPIN, Dipartimento di Scienze Fisiche,
Universit\`a di Napoli Federico II, I-80126, Napoli, Italy
}
\date{\today}% It is always \today, today,

\begin{abstract}
Long-wavelength Mermin-Wagner fluctuations prevent the existence of translational long-range order, in two-dimensional systems at finite temperature. Their dynamical signature, which is the divergence of the vibrational amplitude with the system size, also affects disordered solids and washes out the transient solid-like response generally exhibited by liquids cooled below their melting temperature.
Through a combined numerical and experimental investigation, here we show that long-wavelength fluctuations are also relevant at high temperature, where the liquid dynamics does not reveal a transient solid-like response. In this regime, they induce an unusual but ubiquitous decoupling between long-time diffusion coefficient $D$ and structural relaxation time $\tau$, where $D\propto \tau^{-\kappa}$, with $\kappa > 1$.
Long-wavelength fluctuations have a negligible influence on the relaxation dynamics only at extremely high temperatures, in molecular liquids, or extremely low densities, in colloidal systems.
\end{abstract}

\maketitle

\section{Introduction}

The dimensionality of a system strongly influences the equilibrium properties of its solid phase~\cite{chaikin_lubensky_1995}. According to the Mermin and Wagner theorem~\cite{Mermin}, indeed, systems with continuous symmetry and short-range interactions lack true long-range translational order at finite temperature, in d $\leq$ 2 dimensions. This occurs as in small spatial dimensions the elastic response is dominated by the Goldstone modes, elastic excitations that in the limit of long-wavelength (LW) have a vanishing energy and a diverging amplitude.
A signature of these LW fluctuations is a system size dependent dynamics, which arises as the system size provides a cutoff for the maximum wavelength.
This dependence occurs in both ordered and disordered solids, as LW fluctuations are insensitive to the local order~\cite{Keim_MW}.
This dynamical signature of the LW fluctuations also appears in supercooled liquids, which are liquids cooled below their melting temperature without crystallization occurring.
In particular, LW fluctuations affect the transient solid-like response observed in the supercooled regime, which we recap in Fig.~\ref{fig:dyn} via the investigation of the mean square displacement (MSD), $\Delta r^2(t)$, and of the self-intermediate scattering function (ISF), $F_s(q,t)$, of the two-dimensional (2d) modified Kob-Andersen (mKA) system, a prototypical model glass former (see \emph{Materials and Methods}).
Notice that in the supercooled regime the ISF and the MSD develop a plateau revealing a solid-like response in which particles vibrate in cages formed by their neighbors~\cite{Debenedetti2001}.
%around their equilibrium positions.
However, this is a transient response as at longer times the ISF relaxes and the MSD enters a diffusive behavior.
Flenner and Szamel~\cite{Szamel_2d3d} have demonstrated that this glass relaxation dynamics depends on the system size, the signatures of a transiet solid-like response disappearing in the thermodynamic limit. This trend can be appreciated in Fig.~\ref{fig:dyn}, where we compare the dynamics for two different system sizes, but see Ref.~\cite{Szamel_2d3d} for a full account.
Following works have then demonstrated that this size dependence results from the LW fluctuations~\cite{Keim_MW,Weeks_longwave} by showing that the glassy features of the relaxation dynamics are recovered when the effect of LW fluctuations is filtered out~\cite{Shiba, Kawasaki_LongWL,Weeks_longwave,Keim_MW}.

A transient solid-like response is only observed below the onset temperature, where the relaxation time exhibits a super-Arrhenius temperature dependence, in fragile systems (see \emph{SI Appendix} Fig. S2), and dynamical heterogeneities (DHs) affect the relation between diffusion coefficient and relaxation time~\cite{Debenedetti2001}. In the normal liquid regime that occurs at higher temperatures in molecular liquids, or at lower densities in colloidal systems, the MSD and the ISF do not exhibit plateaus possibly associated to a transient particle localization, and are system size independent. This is apparent in Fig~\ref{fig:dyn}. This suggests that the normal liquid regime is not affected by LW fluctuations. Is this true? And more generally, how far a system should be from the solid phase for the LW fluctuations to have a negligible influence on the relaxation dynamics?
Here we show that, surprisingly, LW fluctuations affect the  structural relaxation dynamics of 2d systems in their normal liquid regime. Specifically, they induce DHs, qualitatively distinct from that observed in the supercooled regime, and an unusual decoupling between the structural relaxation time $\tau$ and the long-time diffusion coefficient, $D$ ({\it Materials and Methods}). This decoupling has been previously observed, both in experiments~\cite{SE_Ellipses} of colloidal systems as well as in numerical simulations~\cite{Sastry_SE,Sung_Tracer}, but its physical origin remained mysterious.
Our results are based on the numerical investigation of the relaxation dynamics of two model glass-forming liquids and on the experimental study of a quasi-2d suspensions of ellipsoids~\cite{SE_Ellipses}.
Numerically, we consider the three-dimensional (3d) Kob-Andersen binary mixture (KA)~\cite{KA_94} and its 2d variant (mKA)~\cite{Kob2009}, as well as the harmonic model~\cite{Corey} (Harmonic) in both 2d and 3d. Numerical details are in {\it Materials and Methods}.
Details on the experimental systems are in Refs.~\cite{Glass_ellipse,SE_Ellipses}.
Our results thus demonstrate that the Mermin-Wagner LW fluctuations
are critical to rationalize the properties of 2d systems not only in the crystalline~\cite{Mermin,2D_melting}, amorphous solid~\cite{Keim_MW} and supercooled phase~\cite{Weeks_longwave}, but also, surprisingly, in the high temperature normal liquid regime.

\section{Results}

\subsection{Heterogeneous dynamics in the 2d normal liquid regime}

\begin{figure}[!t]
\centering
	\includegraphics[angle=0,width=0.4\textwidth]{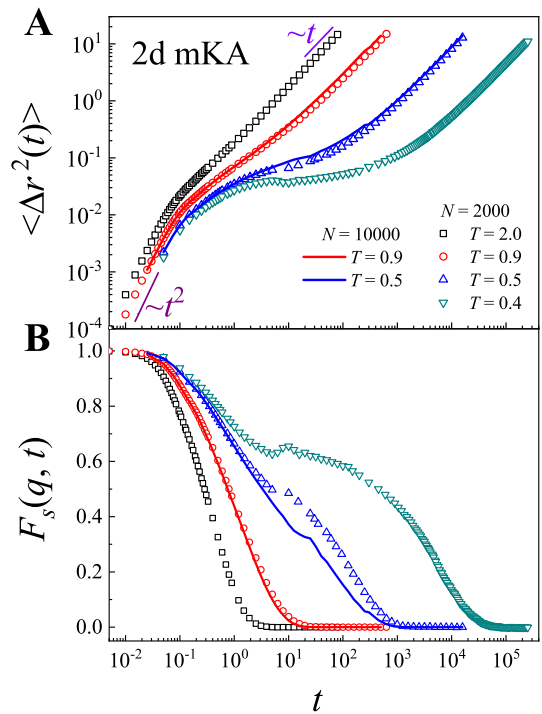}
\caption{($A$) Time dependence of the mean square displacement, and ($B$) of the self-intermediate scattering function from molecular dynamics simulations of the 2d mKA model.
Typical signatures of a supercooled dynamics emerge as the temperature decreases, including ISF oscillations associated to the Boson peak~\cite{Binder2011}.
The comparisons of systems with $N = 2000$ and with $N = 10000$ particles reveal that the dynamics in the supercooled regime is size-dependent~\cite{Szamel_2d3d}, while that in the normal liquid regime is not.
}
\label{fig:dyn}
\end{figure}

\begin{figure}[tb]
\centering
	\includegraphics[angle=0,width=0.45\textwidth]{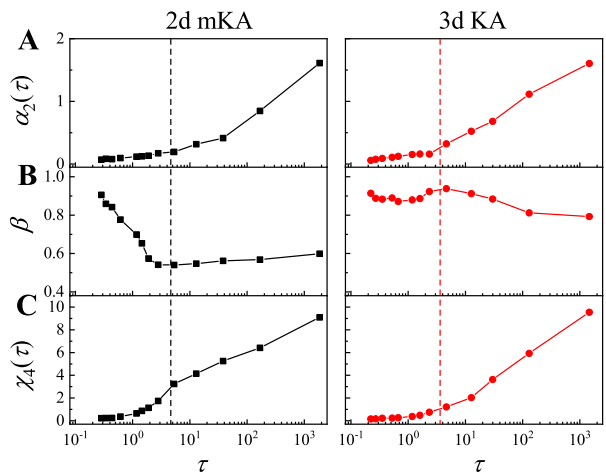}
\caption{
($A$) The non-Gaussian parameter $\alpha_2(\tau)$, ($B$) the stretching exponent $\beta$, and ($C$) the four-point dynamical susceptibility $\chi_4(\tau)$ as functions of relaxation time for 2d mKA model (black squares) and for 3d KA system (red circles). The dashed lines mark the relaxation time at the onset temperature.
\label{fig:intro}
}
\end{figure}

If the long-time relaxation dynamics of a liquid is not influenced by its energy landscape, as suggested by Fig.~\ref{fig:dyn} and by analogous results for the 2d Harmonic model reported in \emph{SI Appendix} Fig. S1, then at the relaxation time scale the displacements of the particles should be uncorrelated, and their van-Hove distribution function a Gaussian.
This is observed in Fig.~\ref{fig:intro}A, where we plot the dependence of the non-Gaussian parameter, a common measure of DHs~\cite{Weeks_Science} (see {\it Materials and Methods}), on the relaxation time, for 2d mKA and the 3d KA models.
Indeed, above the onset temperature, where the relaxation time is that indicated by the dash vertical lines in Fig.~\ref{fig:dyn} (and in the the following figures), $\alpha_2(\tau) \ll 1$, and assumes comparable values in 2d and in 3d.

However, the non-Gaussian parameter is mostly sensitive to deviations from the Gaussian behavior of the tails of the van-Hove distribution, which is populated by the particles with large displacements.
A measure which conversely probes deviations arising from the particles with small displacements can be extracted from the long-time decay of the ISF.
Indeed, if the van-Hove distribution is Gaussian at long time, then the ISF decays exponentially~\cite{Simple_Liquids_book}, while a stretched exponential relaxation $F_s(q,t) \propto \exp(-(t/\tau_e)^\beta)$ with $\beta < 1$ reveals DHs, as commonly observed in the supercooled regime~\cite{Sastry1998}.
Surprisingly, in 2d we observe $\beta < 1$ also in the normal liquid regime, as in Fig.~\ref{fig:intro}B.
Consistently, the dynamical susceptibility at the relaxation time, $\chi_4(\tau)$, which is a proxy for the presence of correlated spatio-temporal motion (see {\it Materials and Methods}), grows in the 2d normal liquid regimes, as in Fig.~\ref{fig:intro}C.
The differences in the behavior of $\beta$ and of $\chi_4(\tau)$ in 2d and in 3d, which are apparent comparing the left and the right column of Fig.~\ref{fig:intro}, signal the presence of DHs in the 2d normal liquid regime.

\begin{figure}[!t]
\centering
	\includegraphics[angle=0,width=0.4\textwidth]{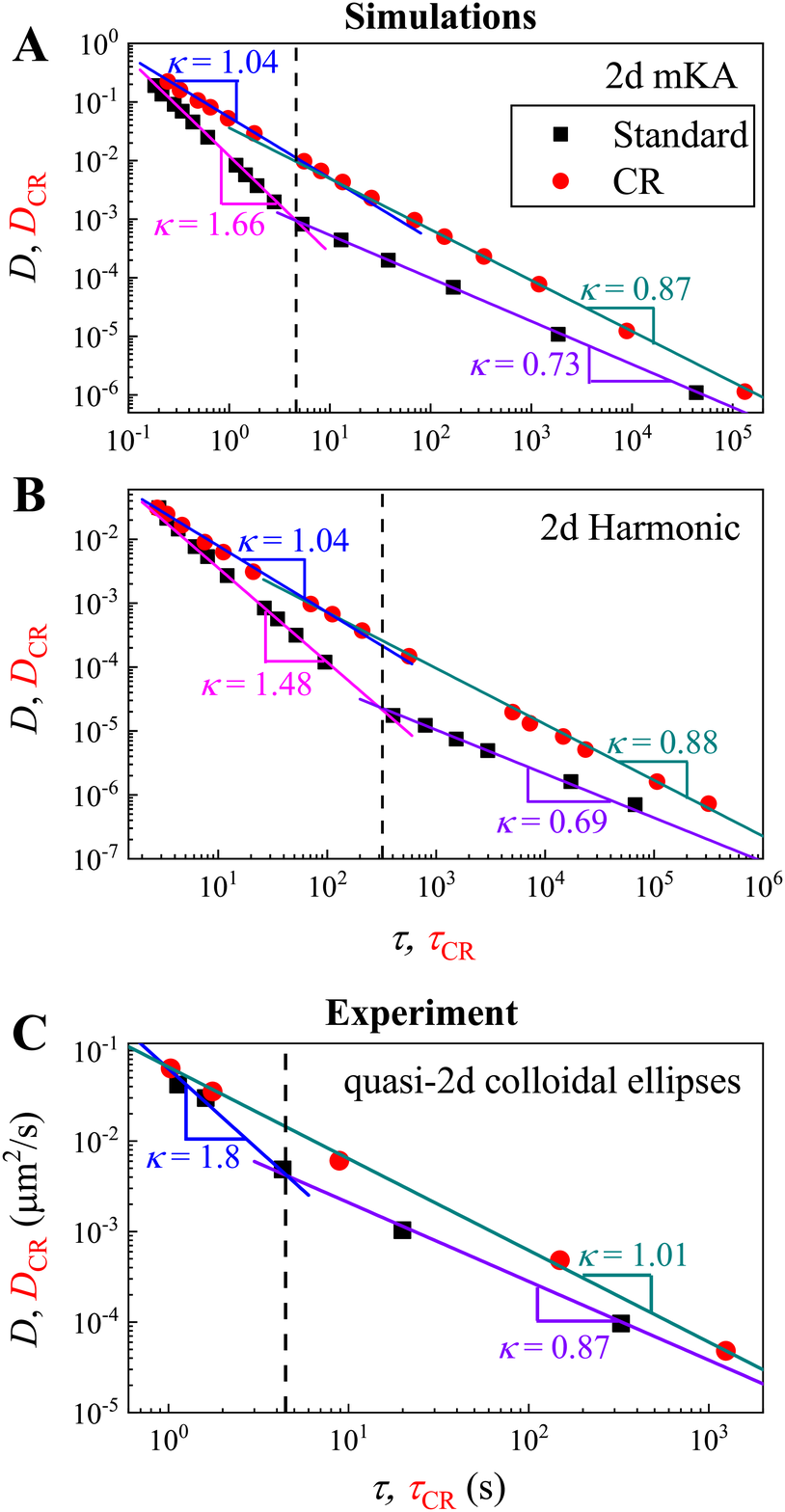}
\caption{
Dependence of the diffusion coefficient $D$ on the relaxation time $\tau$ as obtained investigating the dynamics using the standard (black squares) and the cage-relative (red circles) measures.
Panel $A$ refers to the 2d mKA model, panel $B$ to the 2d Harmonic model and panel $C$ to the experimental quasi-2d colloidal ellipsoids with area fraction $\phi=0.28$, $0.49$, $0.68$, $0.73$, and $0.76$.
\label{fig:Dtau2d}
}
\end{figure}

\subsection{Decoupling between relaxation and diffusion}
The heterogeneities observed in the normal liquid regime of 2d systems are qualitatively different from those observed in the supercooled regime, as they occur in the presence of small values of the non-Gaussian parameter. Hence, these hetereogeneities do not reflect the coexistence of particles with small and large displacements, a hallmark of the supercooled liquid dynamics~\cite{Weeks_Science}.
In supercooled liquids, this coexistence induces a breakdown of the inverse relationship between the diffusion coefficient and the structural relaxation time, leading to $D\propto \tau^{-\kappa}$ with $\kappa < 1$, being $D$ and $\tau$ primarily affected by the particles with the large and small displacements, respectively.
In 2d normal liquids, we therefore not expect to observe $\kappa < 1$.
Indeed, and for reasons that are currently mysterious, one finds $\kappa > 1$~\cite{Sastry_SE, Sung_Tracer, SE_Ellipses}.
We illustrate the crossover from the normal liquid regime where $\kappa > 1$, to the supercooled regime where $\kappa < 1$, in Fig.~\ref{fig:Dtau2d} (black squares). The figure reports results from numerical simulations of the 2d mKA model and of 2d Harmonic discs, as well as results from experiments of quasi-2d colloidal ellipses.
Reader is warned to take with care the precise location of the relaxation time at the onset temperature, dashed line, as well as the value of $\kappa$ in the supercooled regime. Indeed, as we have already seen in Fig.~\ref{fig:intro}, the supercooled regime dynamics has finite-size effects. We will give more details on these size effects later on. Conversely, we stress that results in the normal liquid regime, which is our main focus, are robust.

We provide an insight on the physical origin of the observed heterogeneities, as well as of the $\kappa > 1$ value, by investigating the dynamics using cage-relative (CR) measures (see {\it Materials and Methods}), where the displacement of a particle is evaluated with respect to average displacement of its neighbors.
In Fig.~\ref{fig:Dtau2d} we compare the standard and the CR measures by plotting both $D$ vs. $\tau$ (black squares) and $D_{\rm CR}$ vs. $\tau_{\rm CR}$ (red circles).
In the normal liquid regime we find $\kappa>1$ for the standard measures, and
$\kappa\simeq1$ for the CR measures.
To better illustrate that there exists an extended range of $\tau$ where $\kappa > 1$ for the standard measure,
and $\kappa = 1$ for the CR measures, we show in \emph{SI Appendix} Fig. S3 the presence of a plateau region when the product $\tau_{\rm CR}D_{\rm CR}$ is plotted vs. $\tau_{\rm CR}$.

\begin{figure}[!t]
 \centering
	\includegraphics[angle=0,width=0.48\textwidth]{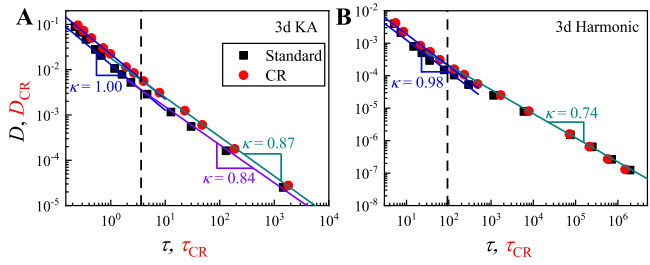}
 \caption{
 The same quantities investigated in Fig.~\ref{fig:Dtau2d} are here studied
 for two 3d systems. Panel A refers to the 3d KA model, while panel B refers
 to the 3d Harmonic model. In 3d $\kappa = 1$ in the normal liquid regime, and
 there is no difference between the standard and the cage-relative measures.
\label{fig:Dtau3d}
}
\end{figure}

These results can be interpreted considering that CR measures have been suggested to filter out the effect of the LW fluctuations on the dynamics~\cite{Weeks_longwave}, in the supercooled regime, as they subtract correlated particle displacements.
Our results suggest that the LW fluctuations are also responsible for the DHs in the normal liquid regime of 2d systems, and for the $\kappa > 1$ value.
In this respect, notice that this large $\kappa$ value indicates that the relaxation time is smaller than expected, given the value of diffusion coefficient, an effect possibly arising from the LW fluctuations.
Indeed, the LW fluctuations have a large amplitude which, when larger than $\frac{2\pi}{q}$, promotes the relaxation of the system but not its diffusion.
This interpretation is supported by the results of Fig.~\ref{fig:Dtau2d}, from which we understand that the standard and the CR measures mainly differ in their estimation of the relaxation time, which is smaller for the standard measure
(see \emph{SI Appendix}, Fig. S4 for a direct comparison of the standard and CR MSD and ISF).
%Fig.~\ref{fig:Dtau2d} also reveal small differences between the standard and the CR diffusivities.
This interpretation is also consistent with the investigation of the difference between the standard and the CR measures in 3d.
Indeed Fig.~\ref{fig:Dtau3d} clearly shows that in 3d, where LW fluctuations play a minor role, there are no noticeable differences between the two measures, neither in the normal liquid nor in the supercooled regime.
We finally notice that Fig.~\ref{fig:Dtau2d} reveals a small difference between the standard and the CR 2d diffusivities, that is not observed in 3d.
This difference cannot originate from the LW fluctuations as it occurs at long-times, as in the diffusive regime there is no elastic response. Rather, they may originate from 2d hydrodynamic fluctuations~\cite{Alder_Hydro,Shiba2019}.

\subsection{Long-wavelength fluctuations in the 2d normal liquid regime}
The above results suggest that in 2d the dynamics of normal liquids is strongly influenced by the LW modes. This is a counter-intuitive speculation as the relaxation dynamics is expected to be weakly dependent on the features of the underlying energy landscape~\cite{Sastry1998,Debenedetti2001} above the onset temperature. To prove the relevance of LW fluctuations in the relaxation dynamics of 2d normal liquid, we therefore perform two additional investigations.

\begin{figure}[!t]
 \centering
	\includegraphics[angle=0,width=0.48\textwidth]{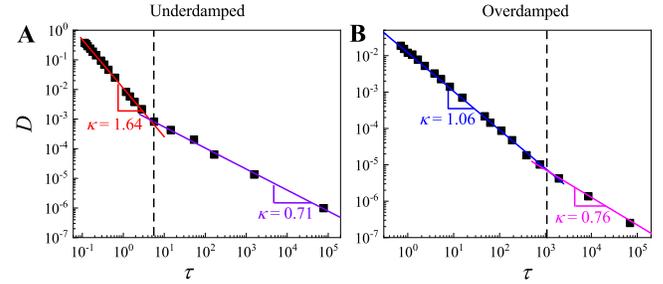}
 \caption{Dependence of the diffusion coefficient $D$ on the relaxation time
$\tau$ for the 2d mKA model. In panel A, the system evolves according to a underdamped dynamics, and $\kappa > 1$ is observed in the normal liquid regime, while in panel B it evolves according to a overdamped dynamics, and $\kappa \simeq 1$.
\label{fig:DtauLD}
}
\end{figure}

First, we consider the effect of the microscopic dynamics on the value of $\kappa$ comparing the two limiting cases of underdamped and overdamped dynamics, corresponding to a Langevin dynamics (see \emph{Materials and Methods}) with Brownian time $\tau_B$ respectively much larger and much smaller than the inverse Debye frequency $1/\omega_D$.
If the LW modes are responsible for the observed behavior, then the value of $\kappa$ must depend on the microscopic dynamics, and $\kappa$ closer to one should be obtained for the overdamped dynamics. Indeed, recall that in the solid phase the contribution of modes with frequency $\omega$ to the MSD is $r_u^2(t,\omega) = \frac{2k_bT}{m\omega^2}[1-\cos(\omega t)]$, in the underdamped limit, and $r_o^2(t,\omega) = \frac{2k_bT}{m\omega^2}[1-e^{-\frac{t}{2\tau_B}}]$ in the overdamped one~\cite{Wang1945}.
Accordingly, while in the overdamped limit the contributions of the different modes have the same time dependence, this is not so in the underdamped limit.
Specifically, in the underdamped limit the modes contribute ballistically to the mean square displacement up to a time $\sim 1/\omega$, which implies that the underdamped dynamics is more strongly affected by the LW modes.
Figure~\ref{fig:DtauLD} compares the relation between $D$ and $\tau$ obtained in the underdamped and in the overdamped limits, for the 2d mKA model, with $\tau_B\omega_D\simeq 100$ and $\simeq 0.01$ respectively, having assumed $\omega_D$ to be of the order of the time at which the ballistic regime ends in the Newtonian dynamics (no damping).
The underdamped results are analogous to those of the previously considered Newtonian dynamics reported in Fig.~\ref{fig:Dtau2d}$A$, as expected. Conversely, the overdamped results strongly differ in that $\kappa \simeq 1$ is essentially recovered in the normal liquid regime.
This agrees with our theoretical argument, and clarifies that the decoupling behavior with $\kappa > 1$ is the combined effect of the presence of LW modes, and of the specific microscopic dynamics through which the system explores its phase space.

\begin{figure}[tb]
\centering
	\includegraphics[angle=0,width=0.48\textwidth]{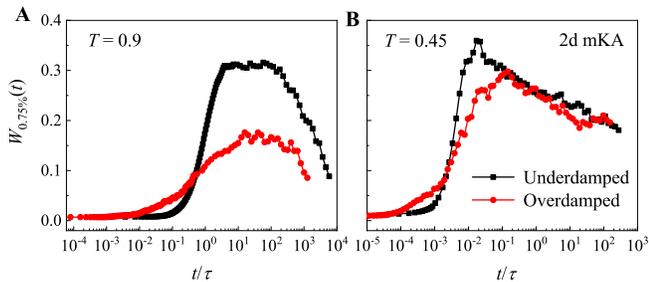}
\caption{
Contribution of the $0.75\%$ of the modes with the longest wavelength to the particle displacement, as a function of time, for system evolving with the underdamped (black squares) and the overdamped (red circles) dynamics.
Panel $A$ is for $T = 0.9$, in the normal liquid regime, while panel $B$ is for $T = 0.45$, in the supercooled regime. The data is for the 2d mKA system, and is obtained averaging over $30$ independent runs.
}
\label{fig:low}
\end{figure}

As a second check we explicitly evaluate the relevance of the LW modes to the overall particle displacement, as a function of time. To this end, we project the normalized displacement $\widehat {\Delta{\mathbf{r}}}(t) = \Delta\mathbf{r}(t)/|\Delta\mathbf{r}(t)|$ of the particles at time $t$ on the eigenvectors $u_i(\omega_i)$ of the Hessian matrix of the nearest inherent configuration (see \emph{Materials and Methods}) of the $t = 0$ configuration: $\widehat{\Delta\mathbf{r}}(t) = \sum_i \beta_i(t) \mathbf{u}_i(\omega_i)$. $\beta^2_i(t)$ is therefore the relative contribution of mode $i$ to the overall displacement at time $t$. To asses the relevance of the LW modes we compute the contribution of the 0.75\% modes with the longest wavelength, $W_{0.75\%}(t) = \sum_i \beta_i^2(t)$.
Figure~\ref{fig:low} compares $W_{0.75\%}(t)$ for the underdamped and the overdamped dynamics, both in the liquid and in the supercooled regime. In both regimes $0.75\%$ of the longest wavelength modes contribute more than $30\%$ of the overall displacement. This is a clear indication that LW fluctuations are extremely relevant, also in the normal liquid regime.
Consistently with our previous finding, we also find the contribution of the LW modes to be more relevant for the underdamped rather than for the overdamped dynamics.
In addition we stress that, in the underdamped regime, for low frequencies $\beta_i(t)$ has a transient oscillatory behavior in line with the presence of transient solid-like modes, also in the liquid regime.
We finally mention that we have verified that these results do not depend on the system size, in the normal liquid regime. Conversely, in the supercooled regime $W_{0.75\%}(t)$ slightly increases with the system size. This system size dependence is consistent with that expected in the presence of LW fluctuations.

\subsection{Stokes-Einstein relation}
\begin{figure}[tb]
 \centering
	\includegraphics[angle=0,width=0.4\textwidth]{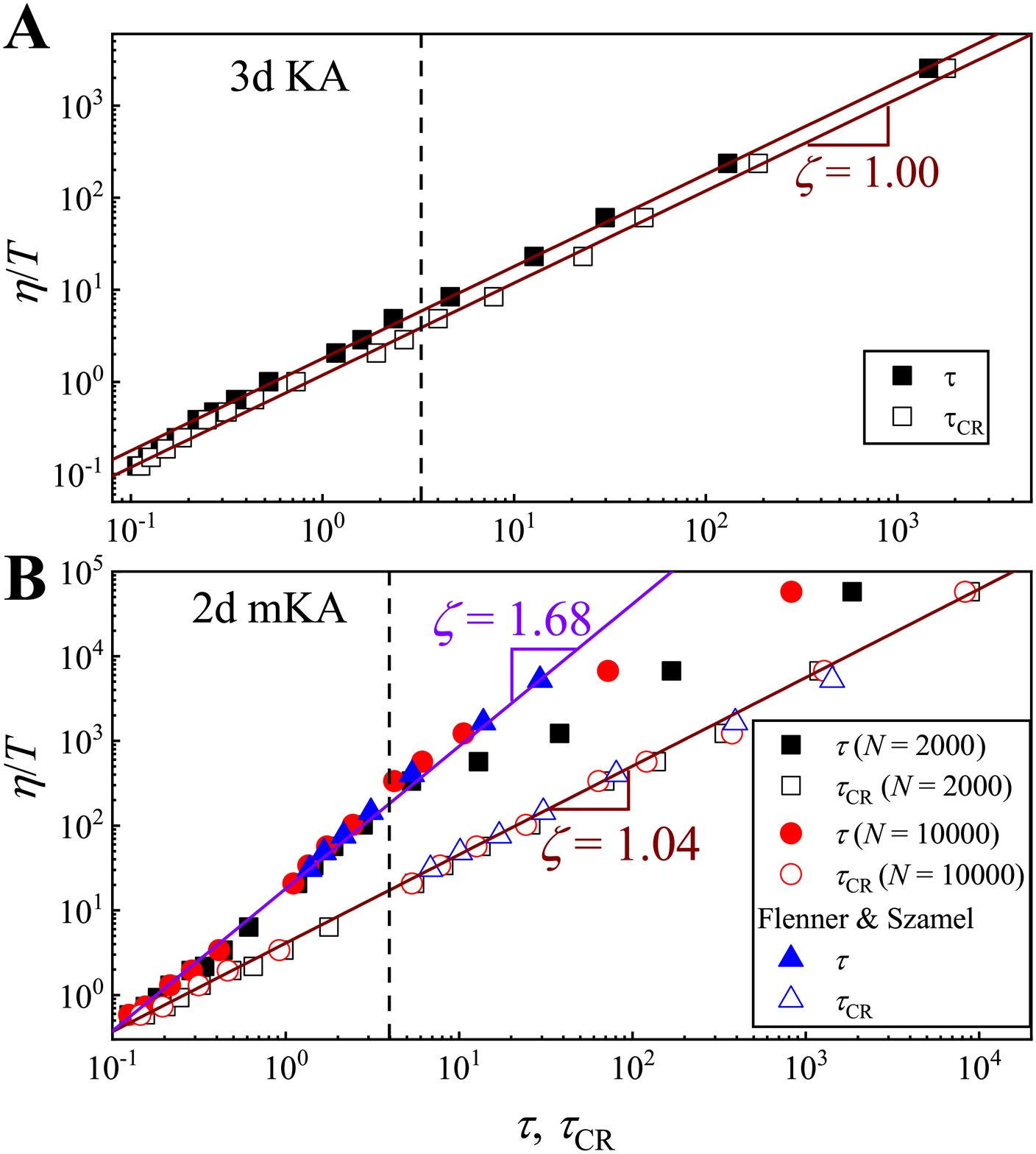}
 \caption{Dependence of the ratio $\eta/T$ on the relaxation time $\tau$ (full symbols) and on the CR relaxation time $\tau_{\rm CR}$ (open symbols). Panel A shows data for the 3d KA model. Panel B illustrates data for the 2d mKA model, and compares different systems sizes using our own data and data from Ref.~\cite{Flenner_2019PNAS}.
\label{fig:eta}
}
\end{figure}

We finally investigate whereas the breakdown of the inverse proportionality between $D$ and $\tau$ implies that of the Stokes-Einstein (SE) relation, $D\propto (\eta/T)^{-1}$, where $\eta$ is the viscosity of the system. This is possible as $\eta$ and $\tau$ are generally related.
To verify this possibility, we measure the viscosity as the Green-Kubo integral~\cite{Simple_Liquids_book} of the shear stress autocorrelation function (see \emph{Materials and Methods}), for both the 3d KA and the 2d mKA model, and study its dependence on the relaxation time.
In 3d, we find $\eta/T \propto \tau \propto \tau_{CR}$, as illustrated in Fig.~\ref{fig:eta}$A$ and previously reported in different systems~\cite{Sastry_SE,Kawasaki_SE}.
In the 2d normal liquid regime $\eta/T\propto\tau^{\zeta}$ with $\zeta=1.68$ and no system size dependence. Since in this regime $D\propto\tau^{-\kappa}$ with $\kappa \simeq \zeta$, as in Fig.~\ref{fig:Dtau2d}$A$, the SE relation holds.
In the supercooled regime we observe a system size dependence, which we highlight also considering data from Ref.~\cite{Flenner_2019PNAS} for systems with up to 4 million particles: the value of the scaling index $\zeta$ increases with the system size, and saturates when the system size is large enough.
The saturation occurs as the time needed for the modes with the longest wavelength to develop, $\propto 1/L$, becomes larger than the relaxation time.
Since the asymptotic value is $\zeta > 1$, while in the supercooled regime $\kappa < 1$ (see Fig.~\ref{fig:Dtau2d}), the SE relation breaks down. Hence, both in 2d and 3d the SE relation only breaks down in the supercooled regime.
In the 2d normal liquid regime the SE does not break down as $\eta$ is determined from the shear stress which, being related to the interparticle distances, is not affected by correlated particle displacements, as the CR measures. Indeed, we show in Fig.~\ref{fig:eta}$B$ (open symbols) that $\eta/T$ is proportional to the CR relaxation time.

\section{Discussion}
In conclusions, our results indicate that LW fluctuations affect the structural relaxation of 2d liquids in the normal liquid regime, where the relaxation dynamics does not suggest a transient solid response, not even when evaluated using CR measures (see \emph{SI Appendix}, Fig. S4).
In this regime the LW fluctuations induce DHs qualitatively different from that observed in the supercooled regime, which is not associated to the coexistence of particles with markedly different displacements. This result allows to rationalize an open issue in the literature~\cite{Sastry_SE,SE_Ellipses,Sung_Tracer}, namely the physical origin of the decoupling between relaxation and diffusion $D\propto \tau^{-\kappa}$ with $\kappa > 1$.
In the main manuscript, we have presented numerical data for the 2d mKA model and the 2d Harmonic model, and their 3d counterpart for comparison, as well as an experimental data for a 2d suspension of hard ellipses. We note, however, that we have also observed consistent results in a binary system with inverse-power-law potential~\cite{Peter_Harrowell} and in monodisperse systems of Penrose kites~\cite{kun,Zong2018,unpublished} (see \emph{SI Appendix}, Fig. S5).
Thus, our findings appear extremely robust as they do not depend on whether the interaction potential is finite or diverging at the origin, attractive or purely repulsive, isotropic or anisotropic.

It is natural to ask if, at high enough temperature or low enough density, the effect of LW fluctuations becomes negligible. The answer to this question is affirmative. Indeed, we do see in Fig.~\ref{fig:Dtau2d} that the difference between the standard and the CR measures, which is a proxy for the relevance of LW fluctuations, decreases as the relaxation time decreases. In the numerical models we have explicitly verified that the two measures coincide in this very high temperature limit. Interestingly, we have found that in this limit the system relaxes before the ballistic regime of the mean square displacement ends. This leads to $D\propto \tau^{-\kappa}$ and $\kappa = 2$, in both 2d and 3d, as we discuss and verify in \emph{SI Appendix}, Fig. S6.
In colloidal systems particles perform independent Brownian motions in the low density limit, where LW fluctuations are therefore negligible. We expect the crossover density below which LW fluctuations are negligible to depend on the viscoelasticity of the solvent.

We conclude with two more remarks.
First, it is established that CR measures remove the effect of LW fluctuations~\cite{Weeks_longwave,Keim_MW}. Here we notice that CR measures filter out all correlated displacements between close particles, regardless of their physical origin. In particular, in the supercooled regime they suppress the effect of correlated particles displacements arising from DHs (see \emph{SI Appendix}, Fig. S7 for the comparison of the four-point dynamical susceptibility between standard and CR measures). This has to be taken into account when using 2d systems to investigate the glass transition.
We notice that it appears difficult to selectively suppress only the correlations arising from one of these two physical processes, as DHs are associated to the low frequency vibrational modes~\cite{Widmer-Cooper2008}. In this respect, perhaps one may consider that DHs in the supercooled regime are associated to localized modes, while LW fluctuations are signatures of extended modes.

Finally, we highlight that LW fluctuations are found in quasi-2d colloidal experiments of both spherical~\cite{Keim_MW,Weeks_longwave} and ellipsoidal~\cite{Zhang2019} particles, as we have shown. However, we have found no clear evidence of LW fluctuations in our overdamped numerical simulations.
Hence, the overdamped simulations do not fully describe the behavior of colloidal suspensions.
This is not a surprise, as it is indeed well known~\cite{Weitz1989,Pine1992,Mason1995,Dhont1996,Raizen2013} that, due to the presence of hydrodynamic interactions, the velocity autocorrelation function of colloidal systems does not decay exponentially as in the numerical simulations of the overdamped dynamics.
The upshot of this consideration is that collective vibrations observed in colloidal systems, including the LW fluctuations, may stem from the hydrodynamic interparticle interaction.
It would be of interest to better characterize these collective hydrodynamic induced modes.

\section{\it Materials and Methods \label{sec:me}}

\subsection{Model systems}
In 2d, we investigated the mKA model~\cite{Kob2009} and the harmonic model~\cite{Corey}.
The mKA model is a mixture of $65\%$ $A$ and $35\%$ $B$ particles. The potential between particles $\alpha$ and $\beta$ is $U_{\alpha\beta}(r) = 4\epsilon_{\alpha\beta}[(\sigma_{\alpha\beta}/r_{\alpha\beta})^{12} - (\sigma_{\alpha\beta}/r_{\alpha\beta})^{6} + C_{\alpha\beta}]$, when $r_{\alpha\beta}$ is smaller than the potential cutoff $r^{c}_{\alpha\beta}$, and zero otherwise. Here, $\alpha, \beta \in$ $\{A, B\}$.
The interaction parameters are given by $\sigma_{AB}/\sigma_{AA}=0.8$, $\sigma_{BB}/\sigma_{AA}=0.88$, $\epsilon_{AB}/\epsilon_{AA}=1.5$, and $\epsilon_{BB}/\epsilon_{AA}=0.5$.
The potential is truncated and shifted at $2.5\sigma_{\alpha\beta}$.  $C_{\alpha\beta}$ guarantees $U_{\alpha\beta}(r^{c}_{\alpha\beta})=0$. The number density is $\rho=1.2$.
Length, energy and time are recorded in units of $\sigma_{AA}$, $\epsilon_{AA}$ and $\sqrt{m\sigma_{AA}^{2}/\epsilon_{AA}}$, respectively. For this model, we consider $N=2000$ (if not otherwise stated) and $N=10000$.
The Harmonic model~\cite{Corey} is a $50:50$ mixture of $N=3000$ particles with interaction potential $U_{\alpha\beta}(r) = 0.5\epsilon(1-r/\sigma_{\alpha\beta})^{2}$, for $r<\sigma_{\alpha\beta}$ and $U_{\alpha\beta}(r) =0$ otherwise.
The size ratios are $\sigma_{AB}/\sigma_{AA}=1.2$ and $\sigma_{BB}/\sigma_{AA}=1.4$, and the number density is $\rho=0.699$. The units for energy, length and time are $\epsilon$, $\sigma_{AA}$, and $\sqrt{m\sigma_{AA}^{2}/\epsilon}$, respectively.
In 3d, we simulated the KA model~\cite{KA_94}, which consists of $N=3074$ with $80\%$ $A$ and $20\%$ $B$ particles, as well as the Harmonic model, with $N = 3000$.
All the results are averaged over at least $4$ independent runs.
We show the data for $A$ particles if the system is a binary mixture.

\subsection{Numerical details}
We have performed simulations in different thermodynamic ensembles,
integrating the equation of motion via a Verlet algorithm.
For the mKA and the KA models, we have performed first simulations in the NVT ensemble to thermalize the system, using a Nos\'{e}-Hoover thermostat, and then in the NVE ensemble for the production runs.
For the mKA model, we have also considered a Langevin dynamics. In this case the equation of motion are $m\mathbf{\ddot{r}}_{i}=-\mathbf{\nabla}_{i} \sum_{j(\neq i)} U(r_{ij})-\gamma \mathbf{\dot{r}}_{i}+\mathbf{\eta}_{i}(t)$, where $\mathbf{r}_{i}$ is the position of the $i$th particle, $r_{ij}$ is the inter-particle distance, $U(r)$ is the potential, $\gamma$ is the friction coefficient and $\mathbf{\eta}_{i}$ is a random noise. The random noise satisfies $\langle\mathbf{\eta}_{i}(t)\mathbf{\eta}_{j}(t')\rangle=2k_{B}T\gamma^{-1}\delta_{ij}
\delta_{t-t'}\mathbf{1}$, with $\mathbf{1}$ the unit tensor.
In the Langevin model, the velocity autocorrelation function decays exponentially with a Brownian timescale $\tau_B = m/\gamma$, in the dilute limit. If $\tau_B$ is the smallest timescale of the problem, this also occurs at finite densities, and the system is overdamped. If $\tau_B$ is very large, conversely, the system is underdamped and behaves as in the NVE ensemble, where the decay of velocity autocorrelation is induced by the interparticle collisions.

For the Harmonic models, we have performed simulations in the NVT+NVE ensemble.
At very low temperature numerical noise causes a small temperature drift in long simulations, in the supercooled regime. In this case the production runs are also carried in the NVT ensemble.

We performed both Newtonian and Langevin dynamics with the GPU-accelerated
GALAMOST package~\cite{Galamost}.
All reported data are taken after equilibrating the system for at least $20\tau$.

Results of Fig.~\ref{fig:low} are obtained by projecting the normalized particle displacement at time $t$ on the modes of the inherent structures of the $t=0$ configurations.
We have obtained these modes minimizing the energy of the $t=0$ configurations using the conjugate gradient method, and then diagonalizing their Hessian matrix.

\subsection{Calculation details}
The MSD is $\langle \Delta r^{2}(t)\rangle=\frac{1}{N}\langle
\sum_{i=1}^{N}\Delta \mathbf{r}_{i}(t)^{2}\rangle$,
where $\Delta \mathbf{r}_{i}(t)=\mathbf{r}_{i}(t)-{\mathbf{r}}_{i}(0)$ is the displacement
of particle $i$ at time $t$. Its long time behavior defines the diffusion coefficient
$D = \lim_{t \to \infty} \frac{\langle r^2(t)\rangle}{2dt}$, with $d$ the dimensionality.
The ISF is $F_{s}(q,t)=1/N\langle
\sum_{j=1}^{N}e^{i\mathbf{q}\cdot\Delta \mathbf{r}_{j}(t)}\rangle$ with
$q=|\mathbf{q}|$ the wavenumber of the first peak of the static structure factor.
The relaxation time $\tau$ is such that $F_{s}(q,\tau) = c$, where $c$ is a small constant. We fix $c = e^{-1}$ as customary, but have checked that this choice does not affect our results.
The non-Gaussian parameter is defined as $\alpha_2(t) = \frac{\langle \Delta x(t)^{4}\rangle}{3\langle \Delta x(t)^{2}\rangle^{2}}-1$ with $ \Delta x(t)$ the displacement in the $x$ coordinate~\cite{Weeks_Science}.
Finally, the four-point dynamical susceptibility $\chi_{4}(t)$ is defined as the variance of the self-intermediate scattering function: $\chi_{4}(t)=N[\langle \hat{F}_{s}(q, t)^{2}\rangle]-\langle \hat{F}_{s}(q, t)\rangle^{2}]$, with $\hat{F}_{s}(q,t)=1/N\sum_{j=1}^{N}e^{i\mathbf{q}\cdot\Delta \mathbf{r}_{j}(t)}$.

The CR quantities, including CR-MSD, CR-ISF and CR four-point susceptibility are defined by replacing the standard displacement with the CR one, which is defined as $\Delta \mathbf{r}_{i}^{\rm CR}(t) =\mathbf{r}_{i}(t)-\mathbf{r}_{i}(0) - 1/N_{j}\sum_{j=1}^{N_{j}}[\mathbf{r}_{j}(t)-\mathbf{r}_{j}(0)]$ with $N_{j}$ the number of neighbors of particle $i$ evaluated at time $0$. We use the Voronoi construction to identify the neighbors of a particle.

The viscosity is computed via the Green-Kubo~\cite{Simple_Liquids_book} integral of the shear stress autocorrelation function, $\eta=\frac{V}{k_{B}T}\int_{0}^{\infty}dt\langle S_{ab}(t)S_{ab}(0)\rangle$, where the stress tensor is  $S_{ab}(t)=\sum_{i=1}^{N}mv_{ia}v_{ib}-\sum_{i=1}^{N}\sum_{j>1}^{N}r_{ija}dU(r_{ij})/dr_{ijb}$,
with $a,b=x,y,z$ indicating the Cartesian components.
To reduce numerical noise, we evaluate the viscosity integrating a numerical fit of the stress autocorrelation function. In the normal liquid regime, $G_{p}/(1+(t/\tau_{p})^{\beta_{p}})$ well describes the stress autocorrelation function, at all times. In the supercooled regime we use a piece-wise approximation, using the above functional form to describe the short time behavior, and a stretched exponential function $G_{e}\exp(-(t/\tau_{e})^{\beta_{e}})$ to describe the long time tails. In any case, the integral is dominated by the long-time decay.

\begin{acknowledgments}
MPC and YWL acknowledge support from the Singapore Ministry of Education
through the Academic Research Fund (Tier 2) MOE2017-T2-1-066 (S) and from the
National Research Foundation Singapore, and are grateful to the National
Supercomputing
Centre (NSCC) of Singapore for providing computational resources.
KZ acknowledges the support from the National Natural Science Foundation of China (21573159 and 21621004).
ZYS acknowledges the support from the National Natural Science Foundation of China (21833008, 21790344).
TGM acknowledges financial support from UCLA.
\end{acknowledgments}

% Bibliography
\bibliography{Maintext}

\begin{thebibliography}{37}%
\makeatletter
\providecommand \@ifxundefined [1]{%
 \@ifx{#1\undefined}
}%
\providecommand \@ifnum [1]{%
 \ifnum #1\expandafter \@firstoftwo
 \else \expandafter \@secondoftwo
 \fi
}%
\providecommand \@ifx [1]{%
 \ifx #1\expandafter \@firstoftwo
 \else \expandafter \@secondoftwo
 \fi
}%
\providecommand \natexlab [1]{#1}%
\providecommand \enquote  [1]{``#1''}%
\providecommand \bibnamefont  [1]{#1}%
\providecommand \bibfnamefont [1]{#1}%
\providecommand \citenamefont [1]{#1}%
\providecommand \href@noop [0]{\@secondoftwo}%
\providecommand \href [0]{\begingroup \@sanitize@url \@href}%
\providecommand \@href[1]{\@@startlink{#1}\@@href}%
\providecommand \@@href[1]{\endgroup#1\@@endlink}%
\providecommand \@sanitize@url [0]{\catcode `\\12\catcode `\$12\catcode
  `\&12\catcode `\#12\catcode `\^12\catcode `\_12\catcode `\%12\relax}%
\providecommand \@@startlink[1]{}%
\providecommand \@@endlink[0]{}%
\providecommand \url  [0]{\begingroup\@sanitize@url \@url }%
\providecommand \@url [1]{\endgroup\@href {#1}{\urlprefix }}%
\providecommand \urlprefix  [0]{URL }%
\providecommand \Eprint [0]{\href }%
\providecommand \doibase [0]{http://dx.doi.org/}%
\providecommand \selectlanguage [0]{\@gobble}%
\providecommand \bibinfo  [0]{\@secondoftwo}%
\providecommand \bibfield  [0]{\@secondoftwo}%
\providecommand \translation [1]{[#1]}%
\providecommand \BibitemOpen [0]{}%
\providecommand \bibitemStop [0]{}%
\providecommand \bibitemNoStop [0]{.\EOS\space}%
\providecommand \EOS [0]{\spacefactor3000\relax}%
\providecommand \BibitemShut  [1]{\csname bibitem#1\endcsname}%
\let\auto@bib@innerbib\@empty
%</preamble>
\bibitem [{\citenamefont {Chaikin}\ and\ \citenamefont
  {Lubensky}(1995)}]{chaikin_lubensky_1995}%
  \BibitemOpen
  \bibfield  {author} {\bibinfo {author} {\bibfnamefont {P.~M.}\ \bibnamefont
  {Chaikin}}\ and\ \bibinfo {author} {\bibfnamefont {T.~C.}\ \bibnamefont
  {Lubensky}},\ }\href {\doibase 10.1017/CBO9780511813467} {\emph {\bibinfo
  {title} {Principles of Condensed Matter Physics}}}\ (\bibinfo  {publisher}
  {Cambridge University Press},\ \bibinfo {year} {1995})\BibitemShut {NoStop}%
\bibitem [{\citenamefont {Mermin}\ and\ \citenamefont {Wagner}(1966)}]{Mermin}%
  \BibitemOpen
  \bibfield  {author} {\bibinfo {author} {\bibfnamefont {N.~D.}\ \bibnamefont
  {Mermin}}\ and\ \bibinfo {author} {\bibfnamefont {H.}~\bibnamefont
  {Wagner}},\ }\href@noop {} {\bibfield  {journal} {\bibinfo  {journal} {Phys.
  Rev. Lett.}\ }\textbf {\bibinfo {volume} {17}},\ \bibinfo {pages} {1133}
  (\bibinfo {year} {1966})}\BibitemShut {NoStop}%
\bibitem [{\citenamefont {Illing}\ \emph {et~al.}(2017)\citenamefont {Illing},
  \citenamefont {Fritschi}, \citenamefont {Kaiser}, \citenamefont {Klix},
  \citenamefont {Maret},\ and\ \citenamefont {Keim}}]{Keim_MW}%
  \BibitemOpen
  \bibfield  {author} {\bibinfo {author} {\bibfnamefont {B.}~\bibnamefont
  {Illing}}, \bibinfo {author} {\bibfnamefont {S.}~\bibnamefont {Fritschi}},
  \bibinfo {author} {\bibfnamefont {H.}~\bibnamefont {Kaiser}}, \bibinfo
  {author} {\bibfnamefont {C.~L.}\ \bibnamefont {Klix}}, \bibinfo {author}
  {\bibfnamefont {G.}~\bibnamefont {Maret}}, \ and\ \bibinfo {author}
  {\bibfnamefont {P.}~\bibnamefont {Keim}},\ }\href@noop {} {\bibfield
  {journal} {\bibinfo  {journal} {Proc. Natl. Acad. Sci. U. S. A.}\ }\textbf
  {\bibinfo {volume} {114}},\ \bibinfo {pages} {1856} (\bibinfo {year}
  {2017})}\BibitemShut {NoStop}%
\bibitem [{\citenamefont {Debenedetti}\ and\ \citenamefont
  {Stillinger}(2001)}]{Debenedetti2001}%
  \BibitemOpen
  \bibfield  {author} {\bibinfo {author} {\bibfnamefont {P.~G.}\ \bibnamefont
  {Debenedetti}}\ and\ \bibinfo {author} {\bibfnamefont {F.~H.}\ \bibnamefont
  {Stillinger}},\ }\href@noop {} {\bibfield  {journal} {\bibinfo  {journal}
  {Nature}\ }\textbf {\bibinfo {volume} {410}},\ \bibinfo {pages} {259}
  (\bibinfo {year} {2001})}\BibitemShut {NoStop}%
\bibitem [{\citenamefont {Flenner}\ and\ \citenamefont
  {Szamel}(2015)}]{Szamel_2d3d}%
  \BibitemOpen
  \bibfield  {author} {\bibinfo {author} {\bibfnamefont {E.}~\bibnamefont
  {Flenner}}\ and\ \bibinfo {author} {\bibfnamefont {G.}~\bibnamefont
  {Szamel}},\ }\href@noop {} {\bibfield  {journal} {\bibinfo  {journal} {Nat.
  Commun.}\ }\textbf {\bibinfo {volume} {6}},\ \bibinfo {pages} {7392}
  (\bibinfo {year} {2015})}\BibitemShut {NoStop}%
\bibitem [{\citenamefont {Vivek}\ \emph {et~al.}(2017)\citenamefont {Vivek},
  \citenamefont {Kelleher}, \citenamefont {Chaikin},\ and\ \citenamefont
  {Weeks}}]{Weeks_longwave}%
  \BibitemOpen
  \bibfield  {author} {\bibinfo {author} {\bibfnamefont {S.}~\bibnamefont
  {Vivek}}, \bibinfo {author} {\bibfnamefont {C.~P.}\ \bibnamefont {Kelleher}},
  \bibinfo {author} {\bibfnamefont {P.~M.}\ \bibnamefont {Chaikin}}, \ and\
  \bibinfo {author} {\bibfnamefont {E.~R.}\ \bibnamefont {Weeks}},\ }\href@noop
  {} {\bibfield  {journal} {\bibinfo  {journal} {Proc. Natl. Acad. Sci. U. S.
  A.}\ }\textbf {\bibinfo {volume} {114}},\ \bibinfo {pages} {1850} (\bibinfo
  {year} {2017})}\BibitemShut {NoStop}%
\bibitem [{\citenamefont {Shiba}\ \emph {et~al.}(2016)\citenamefont {Shiba},
  \citenamefont {Yamada}, \citenamefont {Kawasaki},\ and\ \citenamefont
  {Kim}}]{Shiba}%
  \BibitemOpen
  \bibfield  {author} {\bibinfo {author} {\bibfnamefont {H.}~\bibnamefont
  {Shiba}}, \bibinfo {author} {\bibfnamefont {Y.}~\bibnamefont {Yamada}},
  \bibinfo {author} {\bibfnamefont {T.}~\bibnamefont {Kawasaki}}, \ and\
  \bibinfo {author} {\bibfnamefont {K.}~\bibnamefont {Kim}},\ }\href@noop {}
  {\bibfield  {journal} {\bibinfo  {journal} {Phys. Rev. Lett.}\ }\textbf
  {\bibinfo {volume} {117}},\ \bibinfo {pages} {245701} (\bibinfo {year}
  {2016})}\BibitemShut {NoStop}%
\bibitem [{\citenamefont {Hayato}\ \emph {et~al.}(2018)\citenamefont {Hayato},
  \citenamefont {Peter},\ and\ \citenamefont {Takeshi}}]{Kawasaki_LongWL}%
  \BibitemOpen
  \bibfield  {author} {\bibinfo {author} {\bibfnamefont {S.}~\bibnamefont
  {Hayato}}, \bibinfo {author} {\bibfnamefont {K.}~\bibnamefont {Peter}}, \
  and\ \bibinfo {author} {\bibfnamefont {K.}~\bibnamefont {Takeshi}},\
  }\href@noop {} {\bibfield  {journal} {\bibinfo  {journal} {J. Phys.: Condens.
  Matter}\ }\textbf {\bibinfo {volume} {30}},\ \bibinfo {pages} {094004}
  (\bibinfo {year} {2018})}\BibitemShut {NoStop}%
\bibitem [{\citenamefont {Mishra}\ and\ \citenamefont
  {Ganapathy}(2015)}]{SE_Ellipses}%
  \BibitemOpen
  \bibfield  {author} {\bibinfo {author} {\bibfnamefont {C.~K.}\ \bibnamefont
  {Mishra}}\ and\ \bibinfo {author} {\bibfnamefont {R.}~\bibnamefont
  {Ganapathy}},\ }\href@noop {} {\bibfield  {journal} {\bibinfo  {journal}
  {Phys. Rev. Lett.}\ }\textbf {\bibinfo {volume} {114}},\ \bibinfo {pages}
  {198302} (\bibinfo {year} {2015})}\BibitemShut {NoStop}%
\bibitem [{\citenamefont {Sengupta}\ \emph {et~al.}(2013)\citenamefont
  {Sengupta}, \citenamefont {Karmakar}, \citenamefont {Dasgupta},\ and\
  \citenamefont {Sastry}}]{Sastry_SE}%
  \BibitemOpen
  \bibfield  {author} {\bibinfo {author} {\bibfnamefont {S.}~\bibnamefont
  {Sengupta}}, \bibinfo {author} {\bibfnamefont {S.}~\bibnamefont {Karmakar}},
  \bibinfo {author} {\bibfnamefont {C.}~\bibnamefont {Dasgupta}}, \ and\
  \bibinfo {author} {\bibfnamefont {S.}~\bibnamefont {Sastry}},\ }\href@noop {}
  {\bibfield  {journal} {\bibinfo  {journal} {J. Chem. Phys.}\ }\textbf
  {\bibinfo {volume} {138}},\ \bibinfo {pages} {12A548} (\bibinfo {year}
  {2013})}\BibitemShut {NoStop}%
\bibitem [{\citenamefont {Kim}\ and\ \citenamefont {Sung}(2015)}]{Sung_Tracer}%
  \BibitemOpen
  \bibfield  {author} {\bibinfo {author} {\bibfnamefont {J.}~\bibnamefont
  {Kim}}\ and\ \bibinfo {author} {\bibfnamefont {B.~J.}\ \bibnamefont {Sung}},\
  }\href@noop {} {\bibfield  {journal} {\bibinfo  {journal} {Phys. Rev. Lett.}\
  }\textbf {\bibinfo {volume} {115}},\ \bibinfo {pages} {158302} (\bibinfo
  {year} {2015})}\BibitemShut {NoStop}%
\bibitem [{\citenamefont {Kob}\ and\ \citenamefont {Andersen}(1994)}]{KA_94}%
  \BibitemOpen
  \bibfield  {author} {\bibinfo {author} {\bibfnamefont {W.}~\bibnamefont
  {Kob}}\ and\ \bibinfo {author} {\bibfnamefont {H.~C.}\ \bibnamefont
  {Andersen}},\ }\href@noop {} {\bibfield  {journal} {\bibinfo  {journal}
  {Phys. Rev. Lett.}\ }\textbf {\bibinfo {volume} {73}},\ \bibinfo {pages}
  {1376} (\bibinfo {year} {1994})}\BibitemShut {NoStop}%
\bibitem [{\citenamefont {Brüning}\ \emph {et~al.}(2009)\citenamefont
  {Brüning}, \citenamefont {St-Onge}, \citenamefont {Patterson},\ and\
  \citenamefont {Kob}}]{Kob2009}%
  \BibitemOpen
  \bibfield  {author} {\bibinfo {author} {\bibfnamefont {R.}~\bibnamefont
  {Brüning}}, \bibinfo {author} {\bibfnamefont {D.~A.}\ \bibnamefont
  {St-Onge}}, \bibinfo {author} {\bibfnamefont {S.}~\bibnamefont {Patterson}},
  \ and\ \bibinfo {author} {\bibfnamefont {W.}~\bibnamefont {Kob}},\
  }\href@noop {} {\bibfield  {journal} {\bibinfo  {journal} {J. Phys.: Condens.
  Matter}\ }\textbf {\bibinfo {volume} {21}},\ \bibinfo {pages} {035117}
  (\bibinfo {year} {2009})}\BibitemShut {NoStop}%
\bibitem [{\citenamefont {O’Hern}\ \emph {et~al.}(2003)\citenamefont
  {O’Hern}, \citenamefont {Silbert}, \citenamefont {Liu},\ and\ \citenamefont
  {Nagel}}]{Corey}%
  \BibitemOpen
  \bibfield  {author} {\bibinfo {author} {\bibfnamefont {C.~S.}\ \bibnamefont
  {O’Hern}}, \bibinfo {author} {\bibfnamefont {L.~E.}\ \bibnamefont
  {Silbert}}, \bibinfo {author} {\bibfnamefont {A.~J.}\ \bibnamefont {Liu}}, \
  and\ \bibinfo {author} {\bibfnamefont {S.~R.}\ \bibnamefont {Nagel}},\
  }\href@noop {} {\bibfield  {journal} {\bibinfo  {journal} {Phys. Rev. E}\
  }\textbf {\bibinfo {volume} {68}},\ \bibinfo {pages} {011306} (\bibinfo
  {year} {2003})}\BibitemShut {NoStop}%
\bibitem [{\citenamefont {Mishra}\ \emph {et~al.}(2013)\citenamefont {Mishra},
  \citenamefont {Rangarajan},\ and\ \citenamefont {Ganapathy}}]{Glass_ellipse}%
  \BibitemOpen
  \bibfield  {author} {\bibinfo {author} {\bibfnamefont {C.~K.}\ \bibnamefont
  {Mishra}}, \bibinfo {author} {\bibfnamefont {A.}~\bibnamefont {Rangarajan}},
  \ and\ \bibinfo {author} {\bibfnamefont {R.}~\bibnamefont {Ganapathy}},\
  }\href@noop {} {\bibfield  {journal} {\bibinfo  {journal} {Phys. Rev. Lett.}\
  }\textbf {\bibinfo {volume} {110}},\ \bibinfo {pages} {188301} (\bibinfo
  {year} {2013})}\BibitemShut {NoStop}%
\bibitem [{\citenamefont {Strandburg}(1988)}]{2D_melting}%
  \BibitemOpen
  \bibfield  {author} {\bibinfo {author} {\bibfnamefont {K.~J.}\ \bibnamefont
  {Strandburg}},\ }\href@noop {} {\bibfield  {journal} {\bibinfo  {journal}
  {Rev. Mod. Phys}\ }\textbf {\bibinfo {volume} {60}},\ \bibinfo {pages} {161}
  (\bibinfo {year} {1988})}\BibitemShut {NoStop}%
\bibitem [{\citenamefont {Binder}\ and\ \citenamefont
  {Kob}(2011)}]{Binder2011}%
  \BibitemOpen
  \bibfield  {author} {\bibinfo {author} {\bibfnamefont {K.}~\bibnamefont
  {Binder}}\ and\ \bibinfo {author} {\bibfnamefont {W.}~\bibnamefont {Kob}},\
  }\href {\doibase 10.1142/7300} {\emph {\bibinfo {title} {{Glassy Materials
  and Disordered Solids}}}},\ \bibinfo {edition} {revised}\ ed.\ (\bibinfo
  {publisher} {World Scientific},\ \bibinfo {year} {2011})\BibitemShut
  {NoStop}%
\bibitem [{\citenamefont {Weeks}\ \emph {et~al.}(2000)\citenamefont {Weeks}
  \emph {et~al.}}]{Weeks_Science}%
  \BibitemOpen
  \bibfield  {author} {\bibinfo {author} {\bibfnamefont {E.~R.}\ \bibnamefont
  {Weeks}} \emph {et~al.},\ }\href@noop {} {\bibfield  {journal} {\bibinfo
  {journal} {Science}\ }\textbf {\bibinfo {volume} {287}},\ \bibinfo {pages}
  {627} (\bibinfo {year} {2000})}\BibitemShut {NoStop}%
\bibitem [{\citenamefont {Hansen}\ and\ \citenamefont
  {McDonald}(2005)}]{Simple_Liquids_book}%
  \BibitemOpen
  \bibfield  {author} {\bibinfo {author} {\bibfnamefont {J.~P.}\ \bibnamefont
  {Hansen}}\ and\ \bibinfo {author} {\bibfnamefont {I.~R.}\ \bibnamefont
  {McDonald}},\ }\href@noop {} {\emph {\bibinfo {title} {Theory of Simple
  Liquids, 3rd Ed.}}}\ (\bibinfo  {publisher} {Academic Press, London, UK},\
  \bibinfo {year} {2005})\BibitemShut {NoStop}%
\bibitem [{\citenamefont {Sastry}\ \emph {et~al.}(1998)\citenamefont {Sastry},
  \citenamefont {Debenedetti},\ and\ \citenamefont {Stillinger}}]{Sastry1998}%
  \BibitemOpen
  \bibfield  {author} {\bibinfo {author} {\bibfnamefont {S.}~\bibnamefont
  {Sastry}}, \bibinfo {author} {\bibfnamefont {P.~G.}\ \bibnamefont
  {Debenedetti}}, \ and\ \bibinfo {author} {\bibfnamefont {F.~H.}\ \bibnamefont
  {Stillinger}},\ }\href {\doibase 10.1038/31189} {\bibfield  {journal}
  {\bibinfo  {journal} {Nature}\ }\textbf {\bibinfo {volume} {393}},\ \bibinfo
  {pages} {554} (\bibinfo {year} {1998})}\BibitemShut {NoStop}%
\bibitem [{\citenamefont {Alder}\ and\ \citenamefont
  {Wainwright}(1970)}]{Alder_Hydro}%
  \BibitemOpen
  \bibfield  {author} {\bibinfo {author} {\bibfnamefont {B.~J.}\ \bibnamefont
  {Alder}}\ and\ \bibinfo {author} {\bibfnamefont {T.~E.}\ \bibnamefont
  {Wainwright}},\ }\href@noop {} {\bibfield  {journal} {\bibinfo  {journal}
  {Phys. Rev. A}\ }\textbf {\bibinfo {volume} {1}},\ \bibinfo {pages} {18}
  (\bibinfo {year} {1970})}\BibitemShut {NoStop}%
\bibitem [{\citenamefont {Hayato}\ \emph {et~al.}(2019)\citenamefont {Hayato},
  \citenamefont {Kawasaki},\ and\ \citenamefont {Kim}}]{Shiba2019}%
  \BibitemOpen
  \bibfield  {author} {\bibinfo {author} {\bibfnamefont {S.}~\bibnamefont
  {Hayato}}, \bibinfo {author} {\bibfnamefont {T.}~\bibnamefont {Kawasaki}}, \
  and\ \bibinfo {author} {\bibfnamefont {K.}~\bibnamefont {Kim}},\ }\href@noop
  {} {\bibfield  {journal} {\bibinfo  {journal} {arXiv:1905.05458}\ } (\bibinfo
  {year} {2019})}\BibitemShut {NoStop}%
\bibitem [{\citenamefont {Wang}\ and\ \citenamefont
  {Uhlenbec}(1945)}]{Wang1945}%
  \BibitemOpen
  \bibfield  {author} {\bibinfo {author} {\bibfnamefont {M.~C.}\ \bibnamefont
  {Wang}}\ and\ \bibinfo {author} {\bibfnamefont {G.~E.}\ \bibnamefont
  {Uhlenbec}},\ }\href@noop {} {\bibfield  {journal} {\bibinfo  {journal}
  {Reviews of Modern Physics}\ }\textbf {\bibinfo {volume} {17}},\ \bibinfo
  {pages} {323} (\bibinfo {year} {1945})}\BibitemShut {NoStop}%
\bibitem [{\citenamefont {Kawasaki}\ and\ \citenamefont
  {Kim}(2017)}]{Kawasaki_SE}%
  \BibitemOpen
  \bibfield  {author} {\bibinfo {author} {\bibfnamefont {T.}~\bibnamefont
  {Kawasaki}}\ and\ \bibinfo {author} {\bibfnamefont {K.}~\bibnamefont {Kim}},\
  }\href@noop {} {\bibfield  {journal} {\bibinfo  {journal} {Science Advances}\
  }\textbf {\bibinfo {volume} {3}} (\bibinfo {year} {2017})}\BibitemShut
  {NoStop}%
\bibitem [{\citenamefont {Flenner}\ and\ \citenamefont
  {Szamel}(2019)}]{Flenner_2019PNAS}%
  \BibitemOpen
  \bibfield  {author} {\bibinfo {author} {\bibfnamefont {E.}~\bibnamefont
  {Flenner}}\ and\ \bibinfo {author} {\bibfnamefont {G.}~\bibnamefont
  {Szamel}},\ }\href@noop {} {\bibfield  {journal} {\bibinfo  {journal} {Proc.
  Natl. Acad. Sci. U. S. A.}\ }\textbf {\bibinfo {volume} {116}},\ \bibinfo
  {pages} {2015} (\bibinfo {year} {2019})}\BibitemShut {NoStop}%
\bibitem [{\citenamefont {Perera}\ and\ \citenamefont
  {Harrowell}(1998)}]{Peter_Harrowell}%
  \BibitemOpen
  \bibfield  {author} {\bibinfo {author} {\bibfnamefont {D.~N.}\ \bibnamefont
  {Perera}}\ and\ \bibinfo {author} {\bibfnamefont {P.}~\bibnamefont
  {Harrowell}},\ }\href@noop {} {\bibfield  {journal} {\bibinfo  {journal}
  {Phys. Rev. Lett.}\ }\textbf {\bibinfo {volume} {81}},\ \bibinfo {pages}
  {120} (\bibinfo {year} {1998})}\BibitemShut {NoStop}%
\bibitem [{\citenamefont {Zhao}\ and\ \citenamefont {Mason}(2015)}]{kun}%
  \BibitemOpen
  \bibfield  {author} {\bibinfo {author} {\bibfnamefont {K.}~\bibnamefont
  {Zhao}}\ and\ \bibinfo {author} {\bibfnamefont {T.~G.}\ \bibnamefont
  {Mason}},\ }\href@noop {} {\bibfield  {journal} {\bibinfo  {journal} {Proc.
  Natl. Acad. Sci. U. S. A.}\ }\textbf {\bibinfo {volume} {112}},\ \bibinfo
  {pages} {12063} (\bibinfo {year} {2015})}\BibitemShut {NoStop}%
\bibitem [{\citenamefont {Zong}\ \emph {et~al.}(2018)\citenamefont {Zong},
  \citenamefont {Chen}, \citenamefont {Mason},\ and\ \citenamefont
  {Zhao}}]{Zong2018}%
  \BibitemOpen
  \bibfield  {author} {\bibinfo {author} {\bibfnamefont {Y.}~\bibnamefont
  {Zong}}, \bibinfo {author} {\bibfnamefont {K.}~\bibnamefont {Chen}}, \bibinfo
  {author} {\bibfnamefont {T.~G.}\ \bibnamefont {Mason}}, \ and\ \bibinfo
  {author} {\bibfnamefont {K.}~\bibnamefont {Zhao}},\ }\href {\doibase
  10.1103/PhysRevLett.121.228003} {\bibfield  {journal} {\bibinfo  {journal}
  {Phys Rev Lett}\ }\textbf {\bibinfo {volume} {121}},\ \bibinfo {pages}
  {228003} (\bibinfo {year} {2018})}\BibitemShut {NoStop}%
\bibitem [{\citenamefont {Li}\ \emph {et~al.}(2019)\citenamefont {Li},
  \citenamefont {Li}, \citenamefont {Hou}, \citenamefont {Mason}, \citenamefont
  {Zhao}, \citenamefont {Sun},\ and\ \citenamefont {Ciamarra}}]{unpublished}%
  \BibitemOpen
  \bibfield  {author} {\bibinfo {author} {\bibfnamefont {Y.-W.}\ \bibnamefont
  {Li}}, \bibinfo {author} {\bibfnamefont {Z.-Q.}\ \bibnamefont {Li}}, \bibinfo
  {author} {\bibfnamefont {Z.-L.}\ \bibnamefont {Hou}}, \bibinfo {author}
  {\bibfnamefont {T.~G.}\ \bibnamefont {Mason}}, \bibinfo {author}
  {\bibfnamefont {K.}~\bibnamefont {Zhao}}, \bibinfo {author} {\bibfnamefont
  {Z.-Y.}\ \bibnamefont {Sun}}, \ and\ \bibinfo {author} {\bibfnamefont
  {M.~P.}\ \bibnamefont {Ciamarra}},\ }\href@noop {} {\enquote {\bibinfo
  {title} {Dynamics in two-dimensional glassy systems of crowded penrose
  kites},}\ } (\bibinfo {year} {2019}),\ \bibinfo {note}
  {unpublished}\BibitemShut {NoStop}%
\bibitem [{\citenamefont {Widmer-Cooper}\ \emph {et~al.}(2008)\citenamefont
  {Widmer-Cooper}, \citenamefont {Perry}, \citenamefont {Harrowell},\ and\
  \citenamefont {Reichman}}]{Widmer-Cooper2008}%
  \BibitemOpen
  \bibfield  {author} {\bibinfo {author} {\bibfnamefont {A.}~\bibnamefont
  {Widmer-Cooper}}, \bibinfo {author} {\bibfnamefont {H.}~\bibnamefont
  {Perry}}, \bibinfo {author} {\bibfnamefont {P.}~\bibnamefont {Harrowell}}, \
  and\ \bibinfo {author} {\bibfnamefont {D.~R.}\ \bibnamefont {Reichman}},\
  }\href {\doibase 10.1038/nphys1025} {\bibfield  {journal} {\bibinfo
  {journal} {Nature Physics}\ }\textbf {\bibinfo {volume} {4}},\ \bibinfo
  {pages} {711} (\bibinfo {year} {2008})}\BibitemShut {NoStop}%
\bibitem [{\citenamefont {Zhang}\ and\ \citenamefont
  {Cheng}(2019)}]{Zhang2019}%
  \BibitemOpen
  \bibfield  {author} {\bibinfo {author} {\bibfnamefont {B.}~\bibnamefont
  {Zhang}}\ and\ \bibinfo {author} {\bibfnamefont {X.}~\bibnamefont {Cheng}},\
  }\href@noop {} {\bibfield  {journal} {\bibinfo  {journal} {Soft Matter}\
  }\textbf {\bibinfo {volume} {15}},\ \bibinfo {pages} {4087} (\bibinfo {year}
  {2019})}\BibitemShut {NoStop}%
\bibitem [{\citenamefont {Weitz}\ \emph {et~al.}(1989)\citenamefont {Weitz},
  \citenamefont {Pine}, \citenamefont {Pusey},\ and\ \citenamefont
  {Tough}}]{Weitz1989}%
  \BibitemOpen
  \bibfield  {author} {\bibinfo {author} {\bibfnamefont {D.~A.}\ \bibnamefont
  {Weitz}}, \bibinfo {author} {\bibfnamefont {D.~J.}\ \bibnamefont {Pine}},
  \bibinfo {author} {\bibfnamefont {P.~N.}\ \bibnamefont {Pusey}}, \ and\
  \bibinfo {author} {\bibfnamefont {R.~J.~A.}\ \bibnamefont {Tough}},\ }\href
  {\doibase 10.1103/PhysRevLett.63.1747} {\bibfield  {journal} {\bibinfo
  {journal} {Phys. Rev. Lett.}\ }\textbf {\bibinfo {volume} {63}},\ \bibinfo
  {pages} {1747} (\bibinfo {year} {1989})}\BibitemShut {NoStop}%
\bibitem [{\citenamefont {Zhu}\ \emph {et~al.}(1992)\citenamefont {Zhu},
  \citenamefont {Durian}, \citenamefont {M\"uller}, \citenamefont {Weitz},\
  and\ \citenamefont {Pine}}]{Pine1992}%
  \BibitemOpen
  \bibfield  {author} {\bibinfo {author} {\bibfnamefont {J.~X.}\ \bibnamefont
  {Zhu}}, \bibinfo {author} {\bibfnamefont {D.~J.}\ \bibnamefont {Durian}},
  \bibinfo {author} {\bibfnamefont {J.}~\bibnamefont {M\"uller}}, \bibinfo
  {author} {\bibfnamefont {D.~A.}\ \bibnamefont {Weitz}}, \ and\ \bibinfo
  {author} {\bibfnamefont {D.~J.}\ \bibnamefont {Pine}},\ }\href {\doibase
  10.1103/PhysRevLett.68.2559} {\bibfield  {journal} {\bibinfo  {journal}
  {Phys. Rev. Lett.}\ }\textbf {\bibinfo {volume} {68}},\ \bibinfo {pages}
  {2559} (\bibinfo {year} {1992})}\BibitemShut {NoStop}%
\bibitem [{\citenamefont {Mason}\ and\ \citenamefont
  {Weitz}(1995)}]{Mason1995}%
  \BibitemOpen
  \bibfield  {author} {\bibinfo {author} {\bibfnamefont {T.~G.}\ \bibnamefont
  {Mason}}\ and\ \bibinfo {author} {\bibfnamefont {D.~A.}\ \bibnamefont
  {Weitz}},\ }\href {\doibase 10.1103/PhysRevLett.74.1250} {\bibfield
  {journal} {\bibinfo  {journal} {Phys. Rev. Lett.}\ }\textbf {\bibinfo
  {volume} {74}},\ \bibinfo {pages} {1250} (\bibinfo {year}
  {1995})}\BibitemShut {NoStop}%
\bibitem [{\citenamefont {Dhont}(1996)}]{Dhont1996}%
  \BibitemOpen
  \bibfield  {author} {\bibinfo {author} {\bibfnamefont {J.~K.~G.}\
  \bibnamefont {Dhont}},\ }\href@noop {} {\emph {\bibinfo {title} {{An
  introduction to dynamics of colloids}}}}\ (\bibinfo  {publisher} {Elsevier},\
  \bibinfo {year} {1996})\ p.\ \bibinfo {pages} {642}\BibitemShut {NoStop}%
\bibitem [{\citenamefont {Li}\ and\ \citenamefont {Raizen}(2013)}]{Raizen2013}%
  \BibitemOpen
  \bibfield  {author} {\bibinfo {author} {\bibfnamefont {T.}~\bibnamefont
  {Li}}\ and\ \bibinfo {author} {\bibfnamefont {M.~G.}\ \bibnamefont
  {Raizen}},\ }\href {\doibase 10.1002/andp.201200232} {\bibfield  {journal}
  {\bibinfo  {journal} {Annalen der Physik}\ }\textbf {\bibinfo {volume}
  {525}},\ \bibinfo {pages} {281} (\bibinfo {year} {2013})}\BibitemShut
  {NoStop}%
\bibitem [{\citenamefont {Zhu}\ \emph {et~al.}(2013)\citenamefont {Zhu} \emph
  {et~al.}}]{Galamost}%
  \BibitemOpen
  \bibfield  {author} {\bibinfo {author} {\bibfnamefont {Y.}~\bibnamefont
  {Zhu}} \emph {et~al.},\ }\href@noop {} {\bibfield  {journal} {\bibinfo
  {journal} {J. Comput. Chem.}\ }\textbf {\bibinfo {volume} {34}},\ \bibinfo
  {pages} {2197} (\bibinfo {year} {2013})}\BibitemShut {NoStop}%
\end{thebibliography}%

\end{document}